\documentclass[prd,twocolumn,showpacs,showkeys]{revtex4}

\usepackage{graphicx}
\usepackage{dcolumn}
\usepackage{bm}
\usepackage{amssymb}

\begin{document}

\title{On The Photon Anomalous Magnetic Moment}

\author{S. Villalba Ch\'avez$^{\dag,\ddag}$, H. P\'erez Rojas$^{\ddag}$}

\affiliation{$\dag$ Abdus Salam International Centre for Theoretical Physics, Strada Costiera 11, I-34014, Trieste, Italy.\\
$\ddag$Instituto de Cibernetica, Matematica y Fisica, Calle E 309,
Vedado, Ciudad Habana, Cuba.}

\begin{abstract}
It is shown that due to radiative corrections a photon having a non
vanishing component of its momentum perpendicular to it, bears a
non-zero  magnetic moment. All modes of propagation of the
polarization operator in one loop approximation are discussed and in
this field regime the dispersion equation and the corresponding
magnetic moment are derived. Near the first thresholds of cyclotron
resonance the photon magnetic moment has a peak larger than the
electron anomalous magnetic moment. Related to this magnetic moment,
the arising of some sort of photon "dynamical mass" and a
gyromagnetic ratio are discussed. These latter results might be
interesting in an astrophysical context.
\end{abstract}
\maketitle
\section{Introduction}

It was shown by Schwinger\cite{Schwinger} that electrons get an
anomalous magnetic moment $\mu^\prime=\alpha/2\pi\mu_B$ (being
$\mu_B$ the Bohr magneton) due to QED radiative corrections. By
considering the  propagation of electromagnetic radiation in vacuum
in presence of an external magnetic field, Shabad
\cite{shabad1,shabad2} showed the drastic departure of the photon
dispersion equation from the light cone curve near the energy
thresholds for free pair creation. Three different modes of photon
propagation were found according to the eigenvalues of the
polarization operator in presence of a magnetic field. Some years
later, the same property was obtained \cite{usov1,usov2} in the
vicinity of the threshold for positronium creation. As a result, the
problem of the propagation of light in empty space, in presence of
an external magnetic field is similar to the problem of the
dispersion of light in an anisotropic medium\cite{proceeding}, where
the role of the medium is played by the polarized vacuum in the
external magnetic field. An anisotropy is created by the preferred
direction in space along $\textbf{B}$. In this context we can
mention other characteristics of magnetized quantum vacuum, as the
birefringence\cite{adler,uso3}, (which plays an important role in
the photon splitting and capture effect), and the vacuum
magnetization \cite{Elizabeth}.

From the previous paragraph we may conjecture that, similar to the
case of the electron\cite{Schwinger},  a photon anomalous magnetic
moment might also exist, as a consequence of its radiative
corrections in presence of the magnetic field. We want to show in
what follows that this conjecture is true: the photon having a non
vanishing momentum component orthogonal to a constant magnetic
field, exhibits such anomalous magnetic moment due to  its
interaction  with the virtual electron-positron pairs in the
magnetized vacuum where it propagates.

In order to calculate this photon intrinsic property we recall that
Schwinger's result \cite{Schwinger} was obtained by a weak field
approximation of the electron Green's function in the self energy
operator. In place of using such approximation in the calculation of
the photon anomalous magnetic moment, we prefer to use the exact
form of the polarization operator eigenvalue calculated by Batalin
and Shabad \cite{batalin, shabad1} which provides information about
the three photon propagation modes in the external classical
magnetic field. From it we get the photon equation of motion, in
which we may take the weak as well as the strong field limits. In
any case, we get that the photon acquires a magnetic moment along
the external field, which is proportional to the electron anomalous
magnetic moment.

For the zero field case the photon magnetic moment strictly
vanishes, but for small fields  $\vert\textbf{B}\vert \ll B_c$
(where $B_c=m^2/e\simeq 4.4 \times 10^{13}$G is the Schwinger
critical field) it is significant for a wide range of frequencies.
This might be interesting in an astrophysical context, for instance,
in the propagation and light deflection near a dense magnetized
object.

Moreover, the photon anomalous magnetic moment in the case of strong
magnetic fields, may play an important role in the propagation of
electromagnetic radiation through the magnetospheres of neutron
stars \cite{shabad3} and in other stellar objects where large
magnetic fields arise $\textbf{B}\gtrsim \textbf{B}_c $. The free
and bound pair creation of electrons and positrons at the
thresholds\cite{Hugo1,AO,Leinson}  is related to a production and
propagation of $\gamma$ radiation\cite{denisov,harding}.

The paper has the following structure: in the  Sec. II from the
general solution of the dispersion equation of the polarization
operator  we analyze the interaction term and defined a magnetic
moment of the photon. In section  III we obtain from \cite{shabad1}
the three eigenvalues of the polarization operator calculated in one
loop approximation in the weak field limit and the expression of the
magnetic moment under this field regime.

In Sec. IV and V the magnetic moment is  obtained in the strong
magnetic field limit for the photon and the photon-positronium mixed
state  when both particles are created in the Landau ground state
$n^\prime=n=0$. We will refer also to the case when one of the
particles appear in the first excited state $n=1$, and the other in
the Landau ground state. We discuss mainly the so-called second mode
of propagation near the first pair creation threshold
$n^\prime=n=0$, since as it was studied in \cite{shabad3}, in the
realistic conditions of production and propagation of $\gamma-$
quanta  the third mode decays into the second mode  via the photon
splitting $\gamma\rightarrow\gamma\gamma$ process \cite{adler,uso3}.
Moreover, higher thresholds are damped by the quasi-stationarity of
the electron and positron states on excited Landau levels $n\geq1$
or $n^\prime\geq1$, which may fall down to the ground state with a
photon emitted.

Finally in the Sec. VI the results are analyzed and some remarks and
conclusions are given in Sec. VI on the basis of comparing the
photon behavior with that of a neutral massive particle with
non-vanishing magnetic moment which interacts with the external
field $\vert\bf{B}\vert \simeq B_c$.

\section{The Interaction Energy and The Magnetic Moment of the Radiation}

In paper\cite{batalin} it was shown that the presence of the
constant magnetic field, creates, in addition to the photon momentum
four-vector $k_\mu$, three other four-vectors, $F_{\mu \rho}k^\rho$,
$F^2_{\mu \rho}k^{\rho}$, $F^{*}_{\mu \rho}k^{\rho}$, where $F_{\mu
\nu}=\partial_\mu A_\nu-\partial_\nu A_\mu$ is the electromagnetic
field tensor and $F^*_{\mu \nu}=\frac{i}{2}\epsilon_{\mu \nu \rho
\kappa}F^{\rho \kappa}$ its dual pseudotensor. One get from these
four-vectors three basic independent scalars $k^2$, $kF^2k$,
$kF^{*2}k$, which in addition to the field invariant ${\cal
F}=\frac{1}{4}F_{\mu \rho}F^{\rho \mu}=\frac{1}{2}B^2$, are a set of
four basic scalars of our problem.

In correspondence to each eigenvalue $\pi^{(i)}_{n,n^\prime}$,
$i=1,2,3$, the polarization tensor has  an eigenvector
$a^{(i)}_\mu$(x). The basic vectors \cite{shabad1} are obtained by
normalizing the set of four vectors $C^{1}_\mu= k^2 F^2_{\mu
\lambda}k^\lambda-k_\mu (kF^2 k)$, $C^{2}_\mu=F^{*}_{\mu
\lambda}k^\lambda$, $C^{3}_\mu=F_{\mu \lambda}k^\lambda$,
$C^{4}_\mu=k_\mu$. The first three satisfy in general the
four-dimensional transversality condition $C^{1,2,3}_\mu k^{\mu}=0$,
whereas it is $C^{4}_\mu C^{{4}\mu}=0$ only in the light cone. By
considering $a^{(i)}_\mu (x)$ as the electromagnetic four vector
describing the eigenmodes, it is easy to obtain the corresponding
electric and magnetic fields of each mode ${\bf e}^{(i)}=
\frac{\partial }{\partial x_0}\vec{a}^{(i)}-\frac{\partial
}{\partial {\bf x}}a^{(i)}_0$, ${\bf
h}^{(i)}=\nabla\times\vec{a}^{(i)}$. Up to a factor of
proportionality, we rewrite them from \cite{shabad1}(see also
\cite{Hugo1}),
\begin{eqnarray}
{\bf e}^{(1)}=-\textbf{k}_{\perp} k^2 \omega, \ \ {\bf h}^{(1)}=
[{\bf k}\times {\bf k}_{\parallel} ]k^2\label{wavectors1}, \\
\textbf{e}^{(2)}_{\perp}= {\bf k}_{\perp} k_{\parallel}^2, \ \
\textbf{e}^{(2)}_{\parallel}=\textbf{k}_{\parallel}(k_{\parallel}^2-\omega^2),\
\ \textbf{h}^{(2)}=[{\bf k}_{\perp}\times {\bf k}_{\parallel}]
\omega\label{wavectors2},\\ \textbf{e}^{(3)}=[{\bf k}_{\perp}\times {\bf
k}_{\parallel} ] \omega, \ \  \textbf{h}^{(3)}_{\perp}=-{\bf
k}_{\perp}k_{\parallel}^2,\ \ \textbf{h}^{(3)}_{\parallel}=-{\bf
k}_{\parallel} k_{\perp}^2. \label{wavectors3}
\end{eqnarray}
The previous formulae refer to the reference frame which is at
rest or moving parallel to $B$. The vectors ${\bf k}_{\perp}$ and
${\bf k}_{\parallel}$ are the components of $\textbf{k}$ across
and along $\bf B_z$ (Here  the photon four-momentum squared,
$k^2=k_\perp^2+k_\parallel^2-\omega^2$). It is easy to see that
the mode $i=3$ is a transverse plane polarized wave with its
electric field orthogonal to the plane determined by the vectors
$(\textbf{B}, \textbf{k})$. For propagation orthogonal to $B$, the
mode $a^{(1)}_\mu$ is a pure longitudinal and non physical
electric wave, whereas $a^{(2)}_\mu$ is transverse. For
propagation parallel to $B$, the mode $a^{(2)}_\mu$ becomes purely
electric longitudinal (and non physical), whereas $a^{(1)}_\mu$ is
transverse \cite{shabad1,shabad2}.
In  paper \cite{shabad1} it was  shown that  the dispersion law of
a photon propagating in  vacuum in a strong magnetic field is
given for each mode by the solution of the equations
\begin{equation}
k^2=\pi_{n,n^{\prime}}^{(i)}\left(k^2+\frac{kF^2k}{2\mathcal{F}},\frac{kF^2k}{2\mathcal{F}},B\right),\
\\ \ i=1,2,3 \label{egg}
\end{equation}
The $\pi^\prime$s are the  eigenvalues of the polarization
operator $\Pi_{\mu\nu}(k)$ with the electron and the positron in
the Landau levels $n$ and $n^{\prime}$ or viceversa.

By solving (\ref{egg}) for $z_1=k^2+\frac{kF^2k}{2\mathcal{F}}$ in
terms of $kF^2k/2\mathcal{F}$ it results
\begin{equation}
\omega^2=\vert\textbf{k}\vert^2+f_i\left(\frac{kF^2k}{2\mathcal{F}},B\right)
\label{eg2}
\end{equation}
The term $ f_i\left(kF^2k/2\mathcal{F},B\right)$ is due to the
interaction of the photon with  the virtual $e^{\pm}$ pairs in the
external field, leading to the magnetization of
vacuum\cite{Elizabeth}. Moreover it causes a drastic  departure of
the photon dispersion equation from the light cone curve near the
energy thresholds for free pair creation.

This characteristic  stems from the arising of bound states in the
external field, leading to a singular behavior of the polarization
operator $\Pi_{\mu\nu}(k)$ near the pair creation thresholds  for
electrons and positrons. These particles, coming from the photon
decay in the external field, appear in Landau levels $n$ and
$n^{\prime}$ (cyclotron resonance), or either still stronger
singular behavior of $\Pi_{\mu\nu}$ near the thresholds of an
$e^+e^-$ bound state (due to positronium formation).

To understand the different behavior of $\Pi_{\mu\nu}(k)$ in
presence of an external magnetic field, as compared to the zero
field case, we recall that in the latter problem the polarization
operator is rotationally invariant with regard to the only
significant four-vector, $k_\mu$, whereas in the magnetic field case
this symmetry is reduced to axial. Thus, it is invariant under
rotations in the plane perpendicular to the external field
$\textbf{B}$.

The presence of the interaction energy of the photon with the
electron-positron field opens the possibility of defining a magnetic
moment for the photon, for this we expand the dispersion equation in
linear terms of $\Delta B=B-B_r$ around some field value $B_r$,
\begin{equation}
\label{exp}
\omega(B)=\omega(B_r)+\left.\left(\frac{\partial\omega}{\partial
kF^2k}\frac{\partial kF^2k}{\partial
B}+\frac{\partial\omega}{\partial B}\right)\right\vert_{B=B_r}\Delta
B
\end{equation}

This means that, in the rest frame, where no electric field exists
the photon exhibits  a longitudinal magnetic moment given by
\begin{equation}
\mu_\gamma=-\left.\left(-2k_\perp^2\textbf{B}_z\frac{\partial\omega}{\partial
kF^2k}+\frac{\partial\omega}{\partial
\textbf{B}_z}\right)\right\vert_{B=B_r}\cdot\vec{\kappa}_\parallel\label{treee}
\end{equation}
where $\vec{\kappa}_\parallel$  is an unit vector in the direction
along the magnetic field $B_r=B_z$.

The modulus of the magnetic moment along $\textbf{B}$  can be
expressed as $\mu_\gamma=\mu^\prime g_\gamma$ where the factor
$g_\gamma$ is a sort of gyromagnetic ratio. As in the case of the
electron\cite{lipman} $\mu_{\gamma}$ is not a constant of motion,
but is a quantum average.

As different from the classical theory of  propagation of
electromagnetic radiation  in presence of a constant external
magnetic field, it is expected that $\mu_\gamma$  be different from
zero due to radiative corrections, which are dependent on
$\vert\textbf{B}\vert=B_z=B$. The magnetic moment is induced by the
external field on the photon through its interaction with the
polarized electron-positron virtual quanta of vacuum and is oriented
along $z$.

The gauge invariance property $\pi^{(i)}(0,0)=0$ implies that the
function  $f_i(kF^2k/2\mathcal{F},B)$ vanishes when
$kF^2k/2\mathcal{F}=0$, this means that the anomalous magnetic
moment of the photon is a magnitude subject to the gauge invariance
property of the theory and therefore when the propagation is
parallel to $\bf{B}$, $k_\perp=0$, is cancelled. In every mode,
including positronium formation
\begin{equation}
\mu_\gamma=0\ \ \textrm{if} \ \ k_\perp=0\label{gauge}
\end{equation}
therefore the magnetic moment of the radiation is determined
essentially by the perpendicular photon momentum component and
this determines the optical properties of the quantum vacuum.

Particularly  interesting is the case when $B_r=B_z^r\to0$. If
$\bf{B}$ is assumed small $(\vert\textbf{B}\vert\ll B_c)$, the
dispersion law can be written as
\begin{equation}
\omega=\vert\bf{k}\vert-\mu_\gamma\cdot\bf{B} \label{de0}
\end{equation}
The first term of (\ref{de0}) corresponds to the light cone
equation, whereas the second contains the dipole moment
contribution of the virtual pairs $e^\pm$.

By substitution of (\ref{de0}) in (\ref{wavectors2}) and
(\ref{wavectors3}) we obtain that the electric and magnetic fields
of the radiation corresponding to the second and third modes are
increased by the factors
\begin{equation}
\Delta\textbf{e}^{(2)}_{\parallel}=2\mu_\gamma^{(2)}B_z\vert\textbf{k}\vert\textbf{k}_{\parallel},
\end{equation}
\begin{equation}
\Delta\textbf{h}^{(2)}=\Delta\textbf{e}^{(3)}=-\mu_\gamma^{(2,3)}B_z[{\bf
k}_{\perp}\times {\bf k}_{\parallel}].
\end{equation}
Therefore,  the magnetic moment of the photon leads to linear
effects in quantum electrodynamics and in consequence the refraction
index $n^{(i)}=\vert\textbf{k}\vert/\omega_i$ in mode $i$  is given
by
\begin{equation}
n^{(i)}=1+\frac{\mu_\gamma^{(i)}}{\vert\textbf{k}\vert}
B_z\label{in}
\end{equation}
the gauge invariant property (\ref{gauge}) implies that the
refraction index for parallel propagation, $k_\perp=0$, be exactly
unity: for any mode $n_i=1$. This can be reinterpreted by saying
that for parallel propagation to $B_z$ the refraction index is
equal to unity  due to the vanishing of the photon magnetic
moment.

The components of the group velocity,
($\textbf{v}^{(i)}=\nabla_\textbf{k} \omega_i$),
$\textrm{v}_{\perp\parallel}$ can be written as
 \begin{equation}
\textrm{v}_\perp^{(i)}=\frac{\partial \omega}{\partial
k_\perp}=\frac{k_\perp}{\vert\textbf{k}\vert}\left(1-\frac{\vert\textbf{k}\vert}{k_\perp}\frac{\partial\mu_\gamma^{(i)}}{\partial
k_\perp}B_z\right)\label{vper}
\end{equation}
and
\begin{equation}
\textrm{v}_\parallel^{(i)}=\frac{\partial \omega_i}{\partial
k_\parallel}=\frac{
k_\parallel}{\omega_i}\simeq\frac{k_\parallel}{\vert\textbf{k}\vert}\left(1+\frac{\mu_\gamma^{(i)}}{\vert\textbf{k}\vert}B_z\right).
\label{vpara}
\end{equation}

It follows from (\ref{vper}) and (\ref{vpara}) that  the angle
$\theta^{(i)}$ between the direction of the group velocity and the
external magnetic field  satisfies the relation
\[
\tan\theta^{(i)}=\frac{\textrm{v}_\perp^{(i)}}{\textrm{v}_\parallel^{(i)}}=\left(1-\frac{\vert\textbf{k}\vert}{k_\perp}\frac{\partial\mu_\gamma^{(i)}}{\partial
k_\perp}B_z\right)\left(1+\frac{\mu_\gamma^{(i)}}{\vert\textbf{k}\vert}B_z\right)^{-1}\tan\vartheta,
\] being $\vartheta$ the angle between the photon momentum and
$\textbf{B}$, with $\tan \vartheta=k_\perp/k_\parallel$.

\section{The Polarization Eigenvalues in Weak Field Limit}
In this paper we shall only deal with the transparency region (we do
not consider the absorption of the photon to create observable
$e^{\pm}$ pairs), $\omega^2-k_\parallel^2\leq k_{\perp}^{\prime 2}$
)\textit{i.e.}, we will keep our discussion within the kinematic
domain, where $\pi_{1,2,3}$ are real, where
\begin{equation}
k_{\perp}^{\prime
2}=m_0^2\left[(1+2\frac{B}{B_c}n)^{1/2}+(1+2\frac{B}{B_c}n^{\prime})^{1/2}\right]^2
\end{equation}
is the pair creation squared threshold energy, with the electron and
positron in Landau levels $n,n^{\prime}\neq 0$). We will be
interested in a photon whose energy is near the pair creation
threshold energy.

In the limit $B\ll B_c$ and in one loop approximation the first
and third modes with energies range less than  the first cyclotron
resonance, whose energy is given by $2m_0$, does not show any
singular behavior and in this sense they behave similarly to the
eigenvalues does not contribute. It follows that the dispersion
law for the first and third modes are given by
$\omega_1=\omega_3=\vert\bf{k}\vert$.

Nevertheless, from the calculations made in appendix A it is seen
that the second eigenvalue can be expressed as in the low frequency
limit $4m_0^2\gg k^2+\frac{kF^2k}{\mathcal{F}}$ as
\begin{equation}
\pi_2=-\frac{2\mu^{\prime}B}{m_0}\frac{kF^2k}{2\mathcal{F}}\exp\left(\frac{kF^2k}{4m_0^2\mathcal{F}}\frac{B_c}{B}\right),\label{pi22}\\
\end{equation}
Here $\mu^\prime=(\alpha/2\pi)\mu_B$ is the anomalous magnetic
moment of the electron.

This means that the dispersion equation for the second mode has the
solution
\begin{equation}
\omega^2\simeq
k_\parallel^2-\frac{kF^2k}{2\mathcal{F}}\left(1-\frac{2\mu^{\prime}B}{m_0}\exp\left(\frac{kF^2k}{4m_0^2\mathcal{F}}\frac{B_c}{B}\right)\right)
\label{de1}
\end{equation}

\begin{figure}[!htbp]
\includegraphics[width=3.5in]{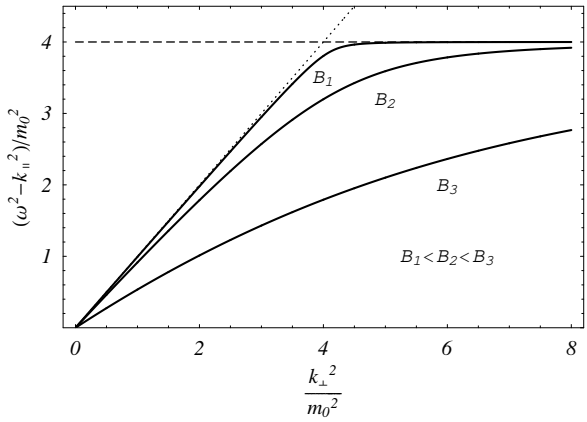}
\caption{\label{fig:edc} Dispersion curves  for second mode  in the
first cyclotron resonance  for different values of weak field. The
dotted line represent the light cone curve, the dashed line
correspond to the threshold energy for pair formation $k_\perp^2 =
4m_0^2$.}
\end{figure}

For values of  energies and magnetic fields for which the
exponential factor in (\ref{pi22}) is of order unity we get
\[
\omega^2=k_\parallel^2-\frac{kF^2k}{2\mathcal{F}}\left(1-\frac{2\mu^{\prime}B}{m_0}\right)
\]
which we can be approximated as
\[
\omega=\vert\textbf{k}\vert+\frac{\mu^{\prime}B}{m_0\vert\textbf{k}\vert}\frac{kF^2k}{2\mathcal{F}}
\]
In such case the  magnetic moment is determined by
\begin{equation}
\mu_\gamma=-\frac{\mu^{\prime}}{m_0\vert\textbf{k}\vert}\frac{kF^2k}{2\mathcal{F}}\label{lemm}
\end{equation}
in the case of perpendicular propagation
\begin{equation}
\mu_\gamma^{max}=\frac{\mu^{\prime}k_\perp}{m_0}
\end{equation}

For perpendicular propagation to $B_z$, $(k_\parallel=0)$, and
although the present approximation is not strictly valid for
$k_\perp\to 2m_0$ the photon anomalous magnetic moment is of order
\begin{equation}
\mu_\gamma \sim 2\mu^{\prime}
\end{equation}

Below we will get a larger value near the first pair creation
threshold. In the table below we show some $\mu_\gamma^{max}$ values
such that $
\exp\left[-\frac{k_\perp^2}{2m_0}\frac{B_c}{B}\right]\sim 1 $
corresponding to  ranges of $X$- rays energies and magnetic field
values.

\begin{widetext}
\begin{tabular}{|c|c|c|c|c|c|c|c|c|}
\cline{1-9} \multicolumn{1}{|c|}{}&
\multicolumn{1}{|c|}{$\omega=m_010^{-8}$} &
\multicolumn{1}{|c|}{$B=1$ G} &
\multicolumn{1}{|c|}{$\omega=m_010^{-6}$} &
\multicolumn{1}{|c|}{$B=10^4$ G} &
\multicolumn{1}{|c|}{$\omega=m_010^{-4}$} &
\multicolumn{1}{|c|}{$B=10^8$ G}&
\multicolumn{1}{|c|}{$\omega=m_010^{-2}$}&
\multicolumn{1}{|c|}{$B=10^{12}$ G}\\
\hline \multicolumn{1}{|c|}{$\mu_{\gamma}^{max}$} &
\multicolumn{2}{|c|}{$10^{-8}\mu^{\prime}$} &
\multicolumn{2}{|c|}{$10^{-6}\mu^{\prime}$} &
\multicolumn{2}{|c|}{$10^{-4}\mu^{\prime}$} &
\multicolumn{2}{|c|}{$10^{-2}\mu^{\prime}$} \\
\hline
\end{tabular}
\\
\\
\end{widetext}

According to (\ref{in}) the refraction index in the weak field
approximation and low frequency limit is given by
\[
n^{(2)}=1+\frac{\mu^\prime k_\perp^2}{m_0\vert\textbf{k}\vert^2}B.
\]

It must be noticed that when the propagation is perpendicular to
$B_z$   the refraction index is maximum
\[
n_\perp^{(2)}=1+\frac{\mu^\prime}{m_0}B
\]

From (\ref{vper}) and (\ref{vpara}) we obtain that the absolute
value  of the group velocity is given by
\begin{equation}
\textrm{v}^{(2)}\simeq
1-\frac{\mu_\gamma^{(2)}}{\vert\textbf{k}\vert}B.\label{vel}
\end{equation}
in the last expression we neglected the  term $B_z$ squared. In the
particular case $k_\parallel=0$
\begin{equation}
\textrm{v}_\perp^{(2)}=1-\frac{\mu^\prime}{m_0}B.
\end{equation}

In the asymptotic region of supercritical magnetic fields $B\gg B_c$
and  restricted energy of longitudinal motion\cite{proceeding}
$\omega^2-k_\parallel^2\ll(B/B_c)m_0^2$, in the low frequency limit
the behavior of the photon propagating in the second mode is equal
to the case of weak magnetic field (\ref{pi22}). When the magnetic
field  $B\sim B_c$, the photon magnetic moment can be approximated
by (\ref{lemm}). For a typical $X$ ray photon of wavelength $\sim$
\AA{}, and propagation perpendicular  to $B_z$, its magnetic moment
has values of order $\mu_\gamma\sim 10^{-2}\mu^\prime$ in this case.
This behavior is also present in the photon-positronium mixed state.

\section{The Cyclotron Resonance}

Similarly to the case of strong magnetic field $(B\gtrsim B_c)$, the
singularity corresponding to cyclotron resonance appears in the
second mode when $B\ll B_c$. The corresponding eigenvalue near of
the first pair creation threshold  is given by
\begin{equation}
\pi_2=\frac{2\alpha m_0^3
B}{B_c}\exp\left(\frac{kF^2k}{4m_0^2\mathcal{F}}\frac{B_c}{B}\right)\left[4m_0^2+\left(k^2+\frac{kF^2k}{2\mathcal{F}}\right)\right]^{1/2},\label{pi2}\\
\end{equation}

The term $\exp(kF^2k B_c/4m_0^2\mathcal{F}B)$ plays an important
role in the cyclotron resonance, both in the weak and the large
field regime. The solution of the equation (\ref{egg}) for the
second mode (first obtained by Shabad \cite{shabad2}) is shown
schematically in Fig \ref{fig:edc}. In this picture, it is noted the
departure of the dispersion law from the light cone curve. For the
first threshold the deflection increases with increasing external
magnetic field. In the vicinity of the first threshold the solutions
of (\ref{egg}) and (\ref{pi2})  are similar to the case of strong
magnetic field regime, therefore, in order to show  the results in a
more compact form let us use the form used in \cite{Hugo2}. In it
the eigenvalues of $\Pi_{\mu\nu}(k\vert A_\mu^{ext})$  near the
thresholds can be written approximately  as
\begin{equation}
\pi_{n,n^{\prime}}^{(i)}\approx-\frac{2\pi\phi_{n,n^{\prime}}^{(i)}}{\vert\Lambda\vert}
\label{eg5}
\end{equation}
with $\vert\Lambda\vert=((k_\perp^{\prime 2 }-k_\perp^{\prime \prime
2})(k_\perp^{\prime 2}+k^2+\frac{kF^2k}{\mathcal{F}}))^{1/2}$
 and
\begin{eqnarray}
k_\perp^{\prime\prime
2}=m_0^2\left[(1+2\frac{B}{B_c}n)^{1/2}-(1+2\frac{B}{B_c}n^{\prime})^{1/2}\right]^2,
\end{eqnarray}
\noindent where $k_{\perp}^{\prime 2}$ is the squared threshold
energy for $e^{\pm}$ pair production, and $k_{\perp}^{\prime\prime
2}$ is the squared threshold energy for excitation between Landau
levels $n,n^{\prime}$ of an electron or positron. The functions
$\phi_{n,n^{\prime}}^{(i)}$ are rewritten from \cite{Hugo1} in the
Appendix B.

In the vicinity of the first resonance $n=n^{\prime}=0$ and
considering $k_\perp\neq0$ and $k_\parallel\neq0$, according to
\cite{shabad1,shabad2} the physical eigenwaves are described by the
second and third modes, but only the second mode has a singular
behavior near the threshold and the function $\phi^{(2)}_{n n'}$ has
the structure
\begin{equation}
\phi_{0,0}^{(2)}\simeq-\frac{2\alpha B m_0^4}{\pi
B_c}\textrm{exp}\left(\frac{kF^2k}{4m_0^2\mathcal{F}}\frac{B_c}{B}\right)
\end{equation}
In this case
 $k_{\perp}^{\prime\prime 2}=0$ and $k_{\perp}^{\prime
2}=4m_0^2$ is the threshold energy.

When $\textbf{B}=\textbf{B}_z$, the scalar
$kF^2k/2\mathcal{F}=-k_\perp^2$ and the approximation of the modes
(\ref{eg5}) turns the dispersion equation (\ref{eg1}) into a cubic
equation in the variable $z_1=\omega^2-k_\parallel^2$ that can be
solved by applying the Cardano formula. We will refer in the
following to (\ref{eg2}) as the real solution of this equation.

We would define the function
\begin{equation}
\Lambda^{*}=(k_\perp^{\prime\prime 2}- k_\perp^{\prime 2})
(k_\perp^2-k_\perp^{\prime 2})
\end{equation}
to simplify the form of the  solutions (\ref{eg2}) of the equation
({\ref{eg1}}). The functions  $f_{i}$ are dependent on
$k_{\perp}^{2},k_{\perp}^{\prime 2},k_{\perp}^{\prime\prime 2}, B$,
and are
\begin{equation}
f_i^{(1)}=\frac{1}{3}\left[2k_\perp^{2}+k_\perp^{\prime
2}+\frac{\Lambda^{* 2}}{(k_\perp^{\prime\prime 2}-k_\perp^{\prime
2})\mathcal{G}^{1/3}}+
\frac{\mathcal{G}^{1/3}}{k_\perp^{\prime\prime 2}-k_{\perp}^{\prime
2}}\right] \label{fi}
\end{equation}
where
\[
\mathcal{G} =6 \pi\sqrt{3}D-\Lambda^{*3}+54 \pi^2
\phi_{n,n^{\prime}}^{(i) 2}(k_\perp^{\prime\prime 2}-k_\perp^{\prime
2})^2
\]
with
\[
D=\sqrt{-(k_\perp^{\prime 2}-k_\perp^{\prime\prime
2})^2\Lambda^{*3}\phi_{n,n^{\prime}}^{(i) 2}\left[1-\frac{27 \pi^2
\phi_{n,n^{\prime}}^{(i)2}(k_\perp^{\prime\prime2}-k_\perp^{\prime
2})^2}{\Lambda^{*3}}\right]}
\]
The  solution $z_1=f_i^{(1)}$, where $z_1=\omega^2-k^2_{\parallel}$,
concern the values of $k_\perp^2$ exceeding the root $k_\perp^{\ast
2}$ of the equation $D=0$, approximately equal to
\[
k_\perp^{\star 2}\simeq k_\perp^{\prime
2}-3\left(\frac{\pi^2\phi_{n,n^{\prime}}^{ (i)2}(k_\perp^{\prime
2})}{k_\perp^{\prime\prime 2}-k_\perp^{\prime 2}}\right)^{1/3}
\]
(for $k_\perp^{2}<k_\perp^{*2}$, $D$ becomes complex). Besides
(\ref{fi}), there are two other solutions of the above- mentioned
cubic equation resulting from substituting (\ref{eg5}) in
(\ref{eg1}). These are complex solutions and are located in the
second sheet of the complex plane of the variable
$z_1=\omega^2-k_\parallel^2$. At $k_\perp^2>k_\perp^{\ast 2}$ these
two complex solutions are given by
\[
f_i^{(2)}=\frac{1}{6}\left[2(2k_\perp^{2}+k_\perp^{\prime
2})-\frac{(1+i\sqrt {3})\Lambda^{* 2}}{(k_\perp^{\prime\prime
2}-k_\perp^{\prime 2})\mathcal{G}^{1/3}}-
\frac{(1-i\sqrt{3})\mathcal{G}^{1/3}}{k_\perp^{\prime\prime
2}-k_\perp^{\prime 2}}\right]
\]
and
\[
f_i^{(3)}=\frac{1}{6}\left[2(2k_\perp^{2}+k_\perp^{\prime
2})-\frac{(1-i\sqrt {3})\Lambda^{* 2}}{(k_\perp^{\prime\prime
2}-k_{\perp}^{\prime 2})\mathcal{G}^{1/3}}-
\frac{(1+i\sqrt{3})\mathcal{G}^{1/3}}{k_\perp^{\prime\prime
2}-k_{\perp}^{\prime 2}}\right]
\]
but they are not interesting to us in the present context.

We should define the functions
\[
m_n=\frac{k_\perp^{\prime}+k_\perp^{\prime\prime }}{2}\ \
\textrm{and}\ \ m_{n^{\prime}}=\frac{k_\perp^{\prime
}-k_\perp^{\prime\prime }}{2},
\]
which are positive for all possible values of $n$ and $n^\prime$.

Now the   magnetic moment of the photon can be derived by taking the
implicit derivative $\partial\omega/\partial B_z$ and
$\partial\omega/\partial kF^2k$ in the dispersion equation, from
(\ref{egg}) and (\ref{eg5})  it is obtained that
\begin{widetext}
\begin{equation}
\mu_\gamma^{(i)}=\frac{\pi}{2\omega(\vert
\Lambda\vert^3-4\pi\phi_{n,n^\prime}^{(i)}m_n
m_{n^\prime})}\left[\phi_{n,n^{\prime}}^{(i)}\left(A\frac{\partial
m_n}{\partial B}+Q\frac{\partial m_{n^\prime}}{\partial
B}\right)-2\Lambda^2\frac{\partial
\phi_{n,n^{\prime}}^{(i)}}{\partial B}\right] \label{mm2}
\end{equation}
being
\[
A=-4m_{n^\prime}[z_1-(m_n+m_{n^\prime})(3m_n+m_{n^\prime})]>0
\]
and
\[
Q=-4m_{n}[z_1-(m_n+m_{n^\prime})(m_n+3m_{n^\prime})]>0.
\]
\end{widetext}

The expression  (\ref{mm2}) contains terms with paramagnetic and
diamagnetic behavior, in which, as is typical in the relativistic
case, the dependence of $\mu$ on $B$ is non-linear. It contains also
diamagnetic terms, depending on the sign of $\phi_{n,
n^{\prime}}^{(i)}$ and its derivatives with regard to $B$: if
$\phi_{n, n^{\prime}}^{(i)}>0$ the first term in the bracket,
contributes paramagnetically ($\frac{\partial m_n}{\partial B}\geq
0$ and $\frac{\partial m_{n^\prime}}{\partial B}\geq 0$), if
$\phi_{n, n^{\prime}}^{(i)}<0$ the sign of this term is opposite and
its contribution is diamagnetic. The second term of (\ref{mm2}) will
be paramagnetic or diamagnetic depending on the sign of the
derivative of the $\phi_{n, n^{\prime}}^{(i)}$ with regard to $B$.

In particular if $n=n^\prime$
\begin{equation}
\mu_\gamma^{(i)}=\frac{\pi}{\omega(\vert
\Lambda\vert^3-4\pi\phi_{n}^{(i)}m_n^2)}\left[\phi_{n}^{(i)}A\frac{\partial
m_n}{\partial B}-\Lambda^2\frac{\partial \phi_{n}^{(i)}}{\partial
B}\right] \label{mm3}
\end{equation}
with $A=-4m_{n}[z_1-8m_n^2]>0$

In the vicinity of the first threshold  $k_\perp^{\prime\prime
2}=0$, $k_\perp^{\prime 2}=4m_0^2$ and $\partial m_n/\partial B=0$
when $n=0$, therefore for the second mode the absolute value of the
magnetic moment is given by
\begin{equation}
\mu_{\gamma}^{(2)}=-\frac{\pi\vert\Lambda\vert^2}{\omega\left(\vert\Lambda\vert^3-4m_0^2\pi\phi_{00}^{(2)}\right)}\frac{\partial
\phi_{00}^{(2)}}{\partial B}\label{FRR}
\end{equation}
One can write the (\ref{FRR}) in explicit form as
\begin{widetext}
\begin{equation}
\mu_\gamma^{(2)}=\frac{\alpha
m_0^3\left(4m_0^2+k^2+\frac{kF^2k}{2\mathcal{F}}\right)\exp\left(\frac{kF^2k}{4m_0^2
\mathcal{F}}\frac{B_c}{B}\right)}{\omega
B_c\left[(4m_0^2+k^2+\frac{kF^2k}{2\mathcal{F}})^{3/2}+\alpha m_0^3
\frac{B}{B_c} \exp\left(\frac{kF^2k}{4m_0^2
\mathcal{F}}\frac{B_c}{B}\right)\right]}\left(1-\frac{kF^2k}{4
m_0^2\mathcal{F}}\frac{B_c}{B}\right).\label{FRR1}
\end{equation}
\end{widetext}
if we consider for simplicity propagation perpendicular to the field
$B$, and $\omega$ near the threshold $2m_0$, the function
$\mu_{\gamma}^{(2)}=f(X)$, where $X=\sqrt{4m_0^2-\omega^2}$ has a
maximum for $X= {\pi\phi_{00}^{(2)}/m_0}^{1/3}$, which is very near
the threshold.

Thus, for perpendicular propagation the expression (\ref{FRR1}) has
a maximum value when $k_\perp^2 \simeq k_\perp^{\prime 2}$.
Therefore in a vicinity of the first pair creation threshold the
magnetic moment of the photon has a paramagnetic behavior and a
resonance peak whose value is given by
\begin{equation}
\mu_{\gamma}^{(2)}=\frac{m_0^2(B+2B_c)}{3m_\gamma
B^2}\left[2\alpha\frac{B}{B_c}\exp\left(-\frac{2B_c}{B}\right)\right]^{2/3}
\label{mumax}
\end{equation}
We would like to define the "dynamical mass" $m_{\gamma}$ of the
photon in presence of a strong magnetic field by the equation
\begin{equation}
m_{\gamma}^{(2)}=\omega (k_\perp^{\prime
2})=\sqrt{4m_0^2-m_0^2\left[2\alpha
\frac{B}{B_c}\exp\left(-\frac{2B_c}{B}\right)\right]^{2/3}}
\label{dm}
\end{equation}

From (\ref{mumax}) it is seen that  near the pair creation
threshold, the magnetic moment induced by the external field in the
photon as a result of its interaction with the polarized
electron-positron quanta of vacuum, has a peak (Fig. 3).  This is a
resonance peak, and we understand it as due to the interaction of
the photon with the polarized $e^{\pm}$ pairs.

The expression (\ref{mumax}) presents a maximum when $B\simeq B_c$,
in such case the magnetic moment is given by
\begin{equation}
\mu_\gamma^{(2)}\approx 3\mu^\prime
\left(\frac{1}{2\alpha}\right)^{1/3}\approx 12.85\mu^\prime
\end{equation}

The arising of a photon dynamical mass is a consequence of the
radiative corrections, which become significant for photon energies
near the pair creation threshold and magnetic fields large enough to
make significant the exponential term $\exp(kF^2k
B_c/4m_0^2\mathcal{F}B) =e^{-k_\perp^2/eB}$. This means that the
massless photon coexists with the massive pair, leading to a
behavior very similar to that of a neutral vector particle
\cite{Osipov} bearing a magnetic moment.One should notice from
(\ref{dm}) that the dynamical mass decrease with the increasing
field intensity, which suggests that the interaction energy of the
photon with its environment increase for magnetic fields far greater
than $B_c$ near the first threshold, when the photon coexist with
the virtual pair.

The  problem of  neutral vector particles is  studied elsewhere
\cite{Herman}. The energy eigenvalues being
\begin{equation}
E(p, B,\eta) =\sqrt{p_3^2+p_{\perp}^2+M^2
+\eta(\sqrt{p_{\perp}^2+M^2}) qB} \label{spec},
\end{equation}
\noindent where the second square root expresses the dependence of
the eigenvalues on the "transverse energy", proportional to the
scalar $pF^2p/{\cal F}$. We have $\eta =0,\pm 1$, and $q$ being a
quantity having the dimensions of magnetic moment. In what follows
we exclude the value $\eta=0$, since it corresponds to the case of
no interaction with the external field $B$. We have that for
$\eta=-1$ the magnetic moment is $\mu= \pm \frac{q}{\sqrt{M^2 -
MqB}}$, which is divergent at the threshold $M =qB$. The behavior of
the photon near the critical field $B_c$ closely resembles this
behavior, since it has a maximum value at the threshold.

For $B_z\gg B_c$ although the vacuum is strongly polarized, the
photon shows a weaker polarization, \emph{i.e} the contribution from
the singular behavior near the thresholds decreases. Actually, the
propagation is being decreased due to an increasing in the imaginary
part of the photon energy $\Gamma$ (we have  the total frequency
$\omega=\omega_r + i\Gamma$, where $\omega_r$ is the real part of
it). As the modes are bent to propagate parallel to $B$, they
propagate in an increasingly absorbent medium.  But for $B>>B_c$ we
are in a region beyond QED and new phenomena related to the standard
model may appear, as for instance, the creation of $\mu^{\pm}$
pairs, and their subsequent decay according to the allowed channels.

As opposite to the second mode case, the polarization eigenvalue
from the third mode in supercritical magnetic field does not
manifest a singular behavior in the first resonance
\cite{proceeding}. In this case the eigenvalues are given by
\begin{widetext}
\begin{equation}
\pi_3=\frac{\alpha
k^2}{3\pi}\left(\ln\frac{B}{B_c}-C\right)+\frac{\alpha}{3\pi}\left(0.21
\frac{kF^2k}{2\mathcal{F}}-1.21\left(k^2+\frac{kF^2k}{2\mathcal{F}}\right)
\right)
\end{equation}
In this case the dispersion equation has the solution
\begin{equation}
\omega^2=\vert\textbf{k}\vert^2+\frac{kF^2k}{2\mathcal{F}}\left[1-\frac{\alpha}{3\pi}\left(\ln\frac{B}{B_c}-C-1.21\right)\right]^{-1}
\label{3m}
\end{equation}
\end{widetext}
being $C$ is Euler constant. For fields $B\sim B_c$ so that the
logarithmic terms are small, the corresponding magnetic moment is
given by
\begin{equation}
\mu_\gamma^{(3)}=\frac{2\mu^{\prime}}{3m_0}\frac{k_\perp^2}{\omega}
\label{3mm}
\end{equation}
for perpendicular propagation to $B_z$ and for photons with energies
near of $m_0$ the magnetic moment of the photon propagating in the
third mode has a value of $\mu_\gamma^{(3)}\sim\mu^{\prime}$.

\section{Magnetic Moment For The Photon-Positronium  Mixed States}

In the case of positronium formation, by following
\cite{shabad3,Leinson}, neglecting the retardation effect and in the
lowest adiabatic approximation,  the Bethe-Salpeter  equation is
reduced  to a Schr\"odinger equation in the variable $(z^e-z^p)$
which  governs  the relative motion along $\textbf{B}$ of the
electron and positron.

Under such approximations, the conservation law induced by the
traslational invariance takes the form $p_x=p_x^p+p_x^e=k_\perp$.
Therefore the binding energy depends on the distance between the
$y$-coordinates of the centers of the electron and positron orbits
$\vert y_0^p-y_0^e\vert=p_x/\sqrt{eB}$.

The  Schr\"odinger equation  mentioned  includes the attractive
coulomb force whose potential in our case has the form
\[
V_{nn^\prime}(z^e-z^p)=-\frac{e^2}{\sqrt{(z^e-z^p)^2+L^4p_x^2}}
\]
where $L=(eB)^{-1/2}$ is the radius of the electron orbit. The
eigenvalue of this equation
$\Delta\varepsilon_{n,n^{\prime}}(n_c,k_\perp^2)$ is the binding
energy of the particles which is numbered by a discrete number
$n_c$ that identifies the Coulomb-bound state for
$\Delta\varepsilon_{n,n^{\prime}}(n_c,k_\perp^2)>0$   and by a
continuous one in the opposite case.

The energy of the pair which does not move along the external
magnetic field is given by
\begin{equation}
\varepsilon_{n,n^\prime}(n_c,k_\perp ^2)=k_\perp^{\prime}+\Delta \varepsilon_{n,n^{\prime}}(n_c,k_\perp^2)
\end{equation}

In this paper we will consider the case in which the Coulomb state
is $n_c=0$, here the binding energy is given by  the expression of
the eigenvalues of this equation,
\begin{equation}
\Delta \varepsilon_{n,n^{\prime}}(0,k_\perp^2)=-\frac{\alpha^2
M_r}{2}\left(2\ln\left[\frac{a_{nn^\prime}^B}{2
\sqrt{L^2+L^4P_x^2}}\right]\right)^{-2},
\end{equation}
where  $a_B^{nn^\prime}=1/e^2M_r$ is Bohr radius and
$M_r=m_nm_{n^\prime}/(m_n+m_{n^\prime})$ the reduced mass of the
bound pair.

The dispersion equation of the positronium is given by
\begin{equation}
k_\perp^2+k_\parallel^2-\omega^2=-\frac{2\pi \Phi_{nn^\prime
n_c}^{(i)}}{\varepsilon_{nn^\prime}^2-\omega^2+k_\parallel^2}
\label{p1}
\end{equation}

For each set of discrete quantum number $n$, $n^\prime$, $n_c$,
equation (\ref{p1}) is quadratic with regard to the variable
$z_1=\omega^2-k_\parallel^2$ and its solutions are
\begin{eqnarray}
f_i=\frac{1}{2}\left(\varepsilon_{nn^\prime
n_c}^2(k_\perp^2)+[k_\perp^2\pm\right.
\\\nonumber\left.(\varepsilon_{nn^\prime
n_c}^2(k_\perp^2)-k_\perp^2)^2-8\pi\Phi_{nn^\prime
n_c}^{(i)}(k_\perp^2)]^{1/2}\right)
\end{eqnarray}

At the first cyclotron resonance $n=n^{\prime}=0$, the function
$\Phi_{000}$ that define the second mode has the structure
\begin{equation}
\Phi_{000}^{(2)}=\phi_{00}^{(2)}\varepsilon_{000}(k_\perp^2)\vert \psi_{000}(0)\vert^2
\end{equation}
with
\[
\Delta\varepsilon_{00}(0,k_\perp^2)=-\alpha^2 m_0\left(\ln\left[\frac{1}{\alpha}\sqrt{\frac{ B_c(1+\frac{k_\perp^2B_c}{m_0^2B})}{B}}\right]\right)^2
\]
where
\[
\vert\psi_{000}(0)\vert^2=\alpha\left\vert\ln\left[\frac{1}{\alpha}\sqrt{\frac{B}{B_c\left(1+\frac{k_\perp^2B_c}{m_0^2B}\right)}}\right]\right\vert
\]
 is the wave function squared of the longitudinal motion\cite{Loudon} at $ z^e=z^p$.

In the case of photon-positronium mixed states we obtain from
(\ref{treee}) and (\ref{p1}) that
\begin{equation}
\mu_{\gamma}^{P}=\frac{\pi\Upsilon}{\omega
(k_\parallel^2-\omega^2-\varepsilon^2)\left
(1+\frac{2\pi\Phi_{nn^\prime
n_c}^{(i)}}{k_\parallel^2-\omega^2-\varepsilon^2} \right)}
\label{mps}
\end{equation}
being
\[
\Upsilon=\varepsilon_{nn^\prime}(n_c) \Phi_{n n^{\prime}
n_c}^{(i)}\frac{\partial\varepsilon}{\partial
B}-\frac{\partial\Phi_{n n^{\prime} n_c}^{(i)}}{\partial B}
\]

Following the  reasoning of (\ref{mumax}), for  magnetic fields
$B\gg B_c$ one can define the dynamical mass of the
photon-positronium mixed state in the first threshold for
positronium energy, $\varepsilon^2 \sim 3.996m_0^2$ and
perpendicular propagation as
\begin{equation}
m_{\gamma}^{P}=\sqrt{\varepsilon_{00}^2-2m_0^2\alpha\left[\frac{B}{B_c}\ln\left(\frac{1}{2\alpha}\frac{B}{B_c}\right)\right]^{1/2}}
\end{equation}
in this regime
\begin{equation}
\mu_{\gamma}^{P}=\frac{m_0^2\alpha\left(1+\ln\left[\frac{B}{2\alpha
B_c}\right]\right)}{2B_cm_\gamma^P\sqrt{\frac{B}{B_c}\ln\left[\frac{B}{2\alpha
B_c}\right]}}
\end{equation}
when $k_\parallel=0$.

In a similar way to the free pair creation, the behavior of the
magnetic moment of the magnetic moment of the mixed state for $n,n'
\neq 0$ can be paramagnetic for some values of Landau numbers and
intervals in momentum space,  and diamagnetic in other ones.

\section{Results and discussion}

\begin{figure}[!htbp]
\includegraphics[width=3in]{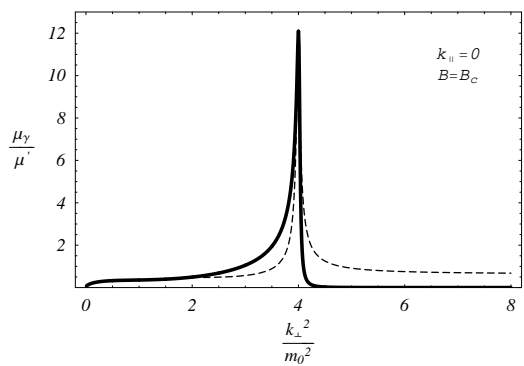}
\caption{\label{fig:Phvk} Magnetic moment curves of the photon
(dark) and photon-positronium mixed states (dashed) with regard to
perpendicular momentum squared,  for the second mode
$k_\perp^{\prime 2}=4m_0^2$ with $n=n^{\prime}=0$.}
\end{figure}

For perpendicular propagation the dependence of the magnetic moment
with regard $k_\perp^2$ in the first resonance $k_\perp^{\prime 2}
=4m_0^2$ is displayed in FIG.  \ref{fig:Phvk} for free pair creation
and photon-positronium mixed state. Both
 curves  show the same qualitative behavior. As it
was expected, near the threshold energy appears the peak
characteristic of the resonance. The result shows that near the pair
creation threshold the magnetic moment of a photon may have values
greater than the anomalous  magnetic moment of the electron.
(Numerical calculations range from 0 to more than $12\mu^\prime$).
This result  is related to the probability of the pair creation,
which is maximum in the first resonant form when $k_\perp=2m_0$, and
increase with $B$, because the medium is becoming absorbent.

We observe that near the thresholds the behavior of the curves is
the same for all  pairs $\omega$, $k_\parallel$ satisfying the
condition $\omega^{2}-k_\parallel^2=k_\perp^{\prime 2}$.

In all curves the magnetic moment decrease for momentum values
$k_\perp^2>4m_0^2$. Therefore the vacuum polarization decreases,
thus, the magnetic moment tends to vanish as it is shown in FIG.
\ref{fig:Phvk}. We interpret these results in the sense that for
photons with squared transversal component of the momenta greater
than $4m_0^2$ the probability of free pair creation in the Landau
ground state (and positronium creation) decrease very fast, since
that region of momenta is to be considered inside the transparency
region corresponding to the next thresholds, i.e., $n=0, n'=1$ or
vice-versa.

\begin{figure}[!htbp]
\includegraphics[width=3in]{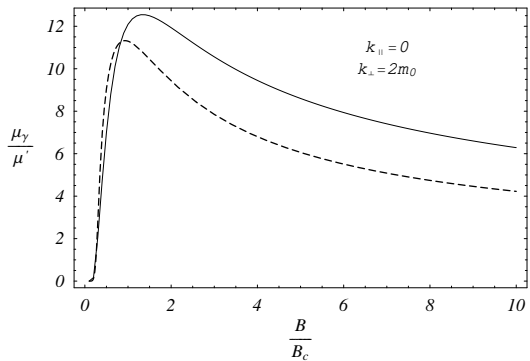}
\caption{\label{fig:mb1} Magnetic moment curves for the photon and
photon-positronium mixed states with regard to external magnetic
field strength. The propagation is perpendicular to $\textbf B$ and
the values of the perpendicular momentum are equal to the absolute
values of the free and bound threshold energies. The first
(continuous) curve refers to the photon whereas the second (dashed)
corresponds to the photon-positronium mixed state.}
\end{figure}

The behavior of the magnetic moment with regard to the field is
shown in the FIG. \ref{fig:mb1} for the case of photon and
photon-positronium mixed state. The picture was obtained in the
field interval $0 \leq B \leq 10B_c$ by considering perpendicular
propagation and taking  the values of the momentum squared as equal
to the absolute values of the threshold energies of free and bound
pair creation. Here, as opposite to the case of low frequency, the
magnetic moment of the photon tend to vanish when $B\rightarrow0$.
We note  that, again, the magnetic moment of the photon is greater
than $\mu^\prime$. In correspondence with figs. \ref{fig:Phvk} the
magnetic moment of the photon not considering Coulomb interaction,
is greater than the corresponding to photon-positronium mixed state.
This result is to be expected  due to the existence of the binding
energy, in the latter case it entails a decrease of the threshold
energy. Each curve has a maximum value, this maximum for the free
pair creation  is approximately $B\approx 1.5 B_c$, whereas for
photon-positron bound state  the value is $B\approx B_c$.

\begin{figure}[!htbp]
\includegraphics[width=3in]{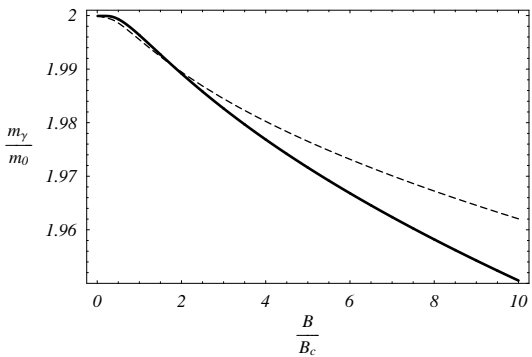}
\caption{\label{fig:mass} Photon dynamical mass dependence with
regard the external  magnetic field of the photon-positronium
(dashed) and photon-free pair (dark) mixed states. Both curves were
obtained near the corresponding first thresholds for each process.}
\end{figure}

In fig. \ref{fig:mass} we display the photon dynamical mass
dependence on the magnetic field, by considering perpendicular
propagation near the thresholds, for free and bound pair creation
and for the second mode. It is shown that for that mode, the
dynamical mass decreases with increasing magnetic field.

For magnetic field values  $B>1.5B_c$ the dynamical mass of the
photon-positronium mixed state is greater than the corresponding
to the case of free pair creation which suggests that the latter
is more probable that the bound state case.

\begin{figure}[!htbp]
\includegraphics[width=3in]{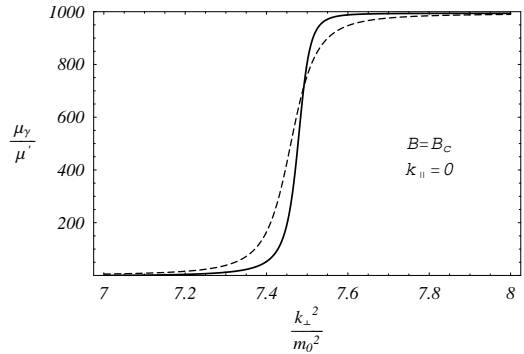}
\caption{\label{su} Curves for the modulus of the photon magnetic
moment with regard to the perpendicular momentum squared, near the
pair creation threshold for the second mode $k_\perp^{\prime 2}=7.46
m_0^2$ with $n=0$, $n^{\prime}=1$ and for perpendicular propagation.
The value of the external magnetic field used for calculation are
$B=B_c$. The dashed line correspond to  photon-positronium mixed
state, whereas the dark line to the free pair creation.}
\end{figure}

\begin{figure}[!htbp]
\includegraphics[width=3in]{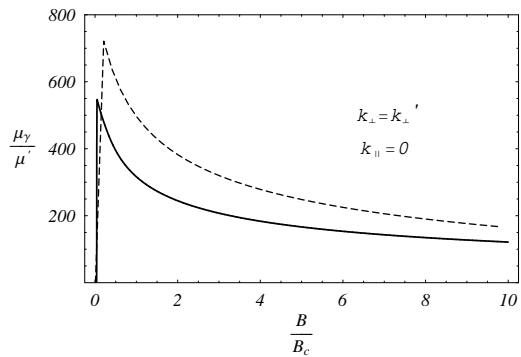}
\caption{\label{suB}Curves for the modulus of the photon magnetic
moment (dark) and photon-positronium mixed states(dashed) plotted
with regard to the external magnetic field strength when the
propagation is perpendicular to $\textbf B$ and the values of the
perpendicular momentum squared is equal to the  values of the free
and bound threshold energies for $n=0$ and $n^\prime=1$.}
\end{figure}

We have found that when the particles are created in excited states
the behavior of the magnetic moment reaches higher
 values with regard the case analyzed previously when
$k_\perp^2=k_\perp^{\prime 2}$ and $B\backsim B_c$. Calculation
points out that these values may be of order $10^{2}\mu^\prime$. The
new values obtained come fundamentally due to the fact of the
threshold energies, which depend on the  magnetic field $\textbf{B}$
(see Fig.2). The new behavior of the photon and photon-positronium
mixed state is shown in the Fig.\ref{su} and Fig. \ref{suB}.

\begin{figure}[!htbp]
\includegraphics[width=3in]{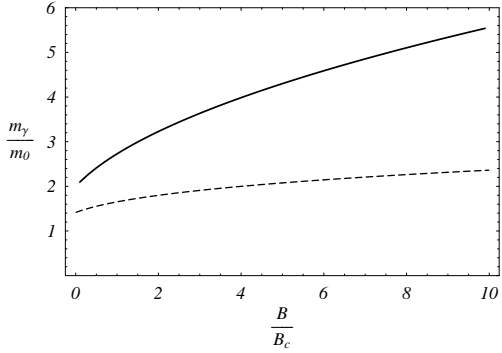}
\caption{\label{m1}Photon dynamical mass dependence with
regard the external  magnetic field of the photon-positronium
(dashed) and photon-free pair (dark)  being $n=0$ and $n^\prime=1$. Both curves
were obtained in the corresponding energy thresholds.}
\end{figure}

In this case the dynamical mass Fig.\ref{m1}, as different from the
previous case, increases with increasing magnetic field. This means
that the magnetic field confine the virtual particles near the
threshold  when these tend to be created in excited states. The
dynamical mass of the positronium in such conditions is always
smaller  than the corresponding to free pair creation, which
suggests that the bound state pair creation in this configuration is
more probable.

\section{Conclusion}

We conclude, first, that a photon propagating in vacuum in presence
of an external magnetic field, exhibits a nonzero magnetic moment
and a sort of dynamical mass due to the magnetic field. This
phenomenon occurs whenever the photon has a nonzero perpendicular
momentum component to the external magnetic field. The values of
this anomalous magnetic moment depend on the propagation mode and
magnetic field regime. The maximum value taken by the photon
magnetic moment is greater than the anomalous magnetic moment of the
electron in a strong magnetic field.

Second, in the small field and low frequency approximations, the
magnetic moment of the photon also exists and is slowly dependent of
the magnetic field intensity in some range of frequencies, whereas
the high frequency limit it depends on $\vert\bf{B}\vert$. In both
cases it vanishes when $B\to0.$

Under these conditions, the behavior of the photon is similar to a
vector neutral massive particle.

\section{Acknowledgement}
Both authors are greatly indebted to Professor A.E. Shabad, from
P. N.Lebedev Physical Institute in Moscow for several comments and
illuminating discussions.

\section{Appendix A}
The three eigenvalues $\pi_i=1,2,3$ of the polarization operator in
one loop approximation, calculated using the exact propagator of
electron in an external magnetic field, can be expressed as linear
combination of three functions $\Sigma_i$. In what follows we will
call $x=B/B_c$
\begin{eqnarray}
&\pi_1&=-\frac{1}{2}k^2\Sigma_1,\\
&\pi_2&=-\frac{1}{2}\left(\left(\frac{kF^2k}{2\mathcal{F}}+k^2\right)\Sigma_2-\frac{kF^2k}{2\mathcal{F}}\Sigma_1\right),\\
&\pi_3&=-\frac{1}{2}\left(\left(\frac{kF^2k}{2\mathcal{F}}+k^2\right)\Sigma_1-\frac{kF^2k}{2\mathcal{F}}\Sigma_3\right).\label{eg1}
\end{eqnarray}
where $\mathcal{F}=\frac{B^2}{2}$ and

We express
\begin{equation}
\Sigma_i=\Sigma_i^{(1)}+\Sigma_i^{(2)},\label{aS}
\end{equation}
being
\begin{equation}
\Sigma_i^{(1)}(x)=\frac{2\alpha}{\pi}\int_0^\infty
dte^{-t/x}\int_{-1}^1d\eta \left[\frac{\sigma_i(t,\eta)}{\sinh
t}-\lim_{t\rightarrow0}\frac{\sigma_i(t,\eta)}{\sinh t}\right]
\end{equation}
and
\begin{widetext}
\begin{eqnarray}
\Sigma_i^{(2)}(x,kF^2k, k^2,
\mathcal{F})=\frac{2\alpha}{\pi}\int_0^\infty
dte^{-t/x}\int_{-1}^1d\eta\frac{\sigma_i(t,\eta)}{\sinh t}
\left[\exp\left(\frac{kF^2k}{2\mathcal{F}}\frac{M(t,\eta)}{m_0^2x}-\left(\frac{kF^2k}{2\mathcal{F}}+k^2\right)\frac{1-\eta^2}{4m_0^2x}t\right)-1\right],
\label{S2}
\end{eqnarray}
\end{widetext}
where
\begin{equation}
M(t,\eta)=\frac{\cosh t -\cosh\eta t}{2\sinh t},
\end{equation}
\begin{equation}
\sigma_1(t,\eta)=\frac{1-\eta }{2}\frac{\sinh(1+\eta)t}{\sinh t},
\label{ss1}
\end{equation}
\begin{equation}
\sigma_2(t,\eta)=\frac{1-\eta^2}{2}\cosh t \label{ss2},
\end{equation}
\begin{equation}
\sigma_3(t,\eta)=\frac{\cosh t -\cosh\eta t}{2\sinh^2 t}.
\label{ss3}
\end{equation}

Express $\Sigma_i^{(1)}$ as
\begin{equation}
\Sigma_i^{(1)}(x)=\frac{2\alpha}{\pi}\int_0^\infty
dte^{-t/x}\left[\frac{g_i(t)}{\sinh t}-\frac{1)}{3t}\right].
\end{equation}
Here
\[
g_i(t)=\int_{-1}^1d\eta\sigma_i(t,\eta)d\eta
\]
and in explicit form
\begin{eqnarray}
&g_1(t)&=\frac{1}{4t\sinh
t}\left(\frac{\sinh2t}{t}-2\right),\label{g1}\\
&g_2(t)&=\frac{\cosh t}{3},\label{g2}\\
&g_3(t)&=\frac{1}{\sinh^2 t}\left(\cosh t-\frac{\sinh
t}{t}\right).\label{g3}
\end{eqnarray}

Let
\begin{equation}
u_i(t)=\frac{g_i(t)}{\sinh t}-\frac{1}{3t}\label{e69}
\end{equation}
The asymptotic expansion of (\ref{e69}) in powers of $\exp(-t)$ have
the form
\begin{equation}
u_1(t)=0, \ \ u_2(t)=1/3,\ \ u_3(t)=0
\end{equation}

Our next purpose is to analyze the  behavior of $\pi_{i}$ when
$x\rightarrow0$. In this case the behavior of $\Sigma_i^{(1)}(x)$ is
determined by the the factor $\exp(-t/x)$ at the integrand which
tends to zero when $x\rightarrow0$. Taking in account the expansion
\[
\exp(-t/x)\simeq \exp(-t/\epsilon)+\frac{\exp(-t/\epsilon
)t}{\epsilon^2}(x-\epsilon),
\]integrating by $t$  and taking
the limit when $\epsilon\rightarrow0$ we obtain that

\begin{eqnarray}
&\Sigma_1^{(1)}&(x)=0,\ \ \Sigma_2^{(1)}(x)\simeq\frac{2\alpha
B}{3\pi B_c},\ \ \Sigma_3^{(1)}(x)=0.\label{ddd}
\end{eqnarray}

The functions $\Sigma_i^{(2)}$  depend of three arguments, as
indicated in (\ref{S2}). The asymptotic expansion of
(\ref{ss1}),(\ref{ss2}), (\ref{ss3}) in powers of $\exp(-t)$ and
$\exp(t\eta)$ produces an expansion of (\ref{S2}) into a sum of
contributions coming from the thresholds, the singular behavior in
the  threshold points originating from the divergencies of the
$t-$integration in ($\ref{S2}$) near $t=\infty$ as it was made in
\cite{shabad1}. The leading term in the expansion of (\ref{ss1}),
(\ref{ss2}), (\ref{ss3}) at $t\rightarrow\infty$ are
\begin{eqnarray}
&\left.\left(\frac{\sigma_1(t,\eta)}{\sinh
t}\right)\right\vert_{t\rightarrow
\infty}&=\frac{1-\eta}{2}\exp\left(-t(1-\eta)\right),\\
&\left.\left(\frac{\sigma_2(t,\eta)}{\sinh
t}\right)\right\vert_{t\rightarrow \infty}&=\frac{1-\eta^2}{4},\\
&\left.\left(\frac{\sigma_3(t,\eta)}{\sinh
t}\right)\right\vert_{t\rightarrow \infty}&=2\exp\left(-2t)\right).
\end{eqnarray}
The function $M(\infty,\eta)=1/2$, one obtains near the lowest
singular threshold ($n=0$, $n^\prime=1$ or viceversa for $i=1$,
$n=n^\prime=1$ for $i=1$, and $n=n^\prime=0$ for $i=2$). Taking into
account this expansion we can write the expression (\ref{S2}) for
the first and their modes as
\begin{widetext}
\begin{equation}
\Sigma_1^{(2)}=\frac{4\alpha e B}{\pi}\int_{-1}^1d\eta(1-\eta)
\left[\frac{\exp\left(\frac{kF^2k}{4m_0^2\mathcal{F}}\frac{B_c}{B}\right)}{4m_0^2+4(1-\eta)eB+\left(k^2+\frac{kF^2k}{2\mathcal{F}}\right)\left(1-\eta^2\right)}-\frac{1}{4
m_0^2+4(1-\eta)eB}\right], \label{SS21}
\end{equation}
\begin{equation}
\Sigma_2^{(2)}=\frac{2\alpha e B}{\pi}\int_{-1}^1d\eta(1-\eta^2)
\exp\left(\frac{kF^2k}{4m_0^2\mathcal{F}}\frac{B_c}{B}\right)\left(\frac{1}{4m_0^2+\left(k^2+\frac{kF^2k}{2\mathcal{F}}\right)\left(1-\eta^2\right)}\right)-\frac{2\alpha
eB}{3\pi m_0^2}, \label{SS22}
\end{equation}
\begin{equation}
\Sigma_3^{(2)}=\frac{16\alpha e B}{\pi}\int_{-1}^1d\eta
\exp\left(\frac{kF^2k}{4m_0^2\mathcal{F}}\frac{B_c}{B}\right)\left[\frac{1}{4m_0^2+8eB+\left(k^2+\frac{kF^2k}{2\mathcal{F}}\right)\left(1-\eta^2\right)}-\frac{1}{4m_0^2+8eB}\right].
\label{SS23}
\end{equation}

In limit $x\to 0$ we get only one expression with singularity in the
first threshold
\begin{equation}
\Sigma_2^{(2)}=\frac{2\alpha B}{3 \pi
B_c}\left(3m_0^2\int_{0}^1d\eta(1-\eta^2)\frac{\exp\left(\frac{kF^2k}{4m_0^2\mathcal{F}}
\frac{B_c}{B}\right)}{4m_0^2+\left(k^2+\frac{kF^2k}{2\mathcal{F}}\right)\left(1-\eta^2\right)}-1\right).\label{As}
\end{equation}

By carrying out the integration on $\eta$ we obtain
\begin{equation}
\Sigma_2^{(2)}=\frac{2\alpha B}{3 \pi
B_c}\left(3m_0^2\exp\left(\frac{kF^2k}{4m_0^2\mathcal{F}}\frac{B_c}{B}\right)
\left[\frac{2}{k^2+\frac{kF^2k}{2\mathcal{F}}}-\frac{8m_0^2\arctan\left(\frac{k^2+\frac{kF^2k}{2\mathcal{F}}}{4m_0^2+k^2+\frac{kF^2k}{2\mathcal{F}}}
\right)^{1/2}}{\left(4m_0^2+k^2+\frac{kF^2k}{2\mathcal{F}}\right)^{1/2}\left(k^2+\frac{kF^2k}{2\mathcal{F}}\right)^{3/2}}\right]-1\right).\label{As0}
\end{equation}

Its behavior in the low frequency limit is given by
\begin{equation}
\Sigma_2^{(2)}=\frac{2\alpha B}{3 \pi
B_c}\left(3m_0^2\exp\left(\frac{kF^2k}{4m_0^2\mathcal{F}}\frac{B_c}{B}\right)\left[\frac{2}{k^2+\frac{kF^2k}{2\mathcal{F}}}-\frac{8m_0^2}{\left(4m_0^2+k^2+\frac{kF^2k}{2\mathcal{F}}\right)\left(k^2+\frac{kF^2k}{2\mathcal{F}}\right)}\right]-1\right),\label{As}
\end{equation}
which we can express as
\begin{equation}
\Sigma_2^{(2)}=\frac{2\alpha B}{3 \pi
B_c}\left(\frac{3m_0^2}{k^2+\frac{kF^2k}{2\mathcal{F}}}\exp\left(\frac{kF^2k}{4m_0^2\mathcal{F}}
\frac{B_c}{B}\right)\left[2-\frac{8m_0^2}{\left(4m_0^2+k^2+\frac{kF^2k}{2\mathcal{F}}\right)}\right]-1\right)\label{As}
\end{equation}
\end{widetext}
therefore
\begin{equation}
\Sigma_2^{(2)}=\frac{2\alpha B}{3 \pi
B_c}\left(\frac{3}{2}\exp\left(\frac{kF^2k}{4m_0^2\mathcal{F}}\frac{B_c}{B}\right)-1\right).\label{As}
\end{equation}

Under such condition, by substituting the last one and the second
expression of (\ref{ddd}) in (\ref{eg1}), we have that the second
eigenvalue of the polarization operator can be written as
\begin{equation}
\pi_2=-\frac{2\mu^{\prime}B}{m_0}\left(\frac{kF^2k}{2\mathcal{F}}+k^2\right)\exp\left(\frac{kF^2k}{4m_0^2\mathcal{F}}\frac{B_c}{B}\right).
\end{equation}
In the same approximation
$\frac{kF^2k}{2\mathcal{F}}+k^2\approx\frac{kF^2k}{2\mathcal{F}}$
and
\begin{equation}
\pi_2=-\frac{2\mu^{\prime}B}{m_0}\frac{kF^2k}{2\mathcal{F}}\exp\left(\frac{kF^2k}{4m_0^2\mathcal{F}}\frac{B_c}{B}\right).
\end{equation}

Now, the behavior  $\pi_2$ near of the first threshold can be
determined from (\ref{As0}) by taking into account  the point
$\omega^2-k_\parallel=4m_0^2-\epsilon$ with $\epsilon>0$ and
$\epsilon\rightarrow0$
\begin{eqnarray}
\pi_2=\frac{2\alpha m_0^3
B}{B_c}\exp\left(\frac{kF^2k}{4m_0^2\mathcal{F}}\frac{B_c}{B}\right)\left[4m_0^2+\left(k^2+\frac{kF^2k}{2\mathcal{F}}\right)\right]^{1/2}.
\end{eqnarray}

\section{Appendix B}
\begin{widetext}
The function $\phi_{n,n^\prime}^{(i)}$ is given by
\begin{equation}
\phi_{n,n^\prime}^{(1)}=-\frac{e^2}{4 \pi^2}\frac{e B
k^2}{z_1}\left[(2eB(n+n^\prime)+z_1)F_{n,n^\prime}^{(2)}-4k_\perp^2N_{n,n^\prime}^{(1)}\right],
\end{equation}
\begin{equation}
\phi_{n,n^\prime}^{(2)}=-\frac{e^2}{4 \pi^2}e B
\left[\left(\frac{2e^2B^2(n-n^\prime)^2}{z_1}+2m^2+eB(n+n^\prime)\right)F_{n,n^\prime}^{(1)}+2eB
(n n^\prime)^{1/2} G_{n,n^\prime}^{(1)}\right],
\end{equation}
\begin{equation}
\phi_{n,n^\prime}^{(3)}=-\frac{e^2}{4 \pi^2}\frac{e B
k^2}{z_1}\left[(2eB(n+n^\prime)+z_1)F_{n,n^\prime}^{(2)}+4k_\perp^2N_{n,n^\prime}^{(1)}\right],
\end{equation}
where, calling $y=\frac{kF^2k}{4m_0^2\mathcal{F}}\frac{B_c}{B}$ and
$z_1=k^2+\frac{kF^2k}{2\mathcal{F}}$
\[
F_{n,n^\prime}^{(1)}=\left\{[L_{n^\prime-1}^{n-n^\prime}(y)]^2+\frac{n^\prime}{n}[L_{n^\prime}^{n-n^\prime}(y)]^2\right\}\frac{(n^\prime-1)!}{(n-1)!}y^{n-n^\prime}\exp[-y],
\]
\[
F_{n,n^\prime}^{(2,3)}=\left\{\frac{y}{n}[L_{n^\prime-1}^{n-n^\prime+1}(y)]^2\pm\frac{n^\prime}{x}[L_{n^\prime}^{n-n^\prime-1}(y)]^2\right\}\frac{(n^\prime-1)!}{(n-1)!}y^{n-n^\prime}\exp[-y],
\]
\[
G_{n,n^\prime}^{(1)}=2\left(\frac{n^\prime}{n}\right)^{1/2}\frac{(n^\prime-1)!}{(n-1)!}y^{n-n^\prime}L_{n^\prime-1}^{n-n^\prime}(y)L_{n^\prime}^{n-n^\prime}(y)\exp[-y],
\]
\[
N_{n,n^\prime}^{(1)}=\frac{n^\prime!}{(n-1)!}y^{n-n^\prime-1}L_{n^\prime-1}^{n-n^\prime+1}(y)L_{n}^{n-n^\prime-1}(y)\exp[-y].
\]

Here $L_n^m(y)$ are generalized Laguerre polynomials. Laguerre
polynomials with $-1$ for lower index must be taken as zero.
\end{widetext}

\bibliographystyle{apsrev}

\end{document}